# Coherence interpretation of the noninterfering Sagnac-based quantum correlation

Byoung S. Ham
School of Electrical Engineering and Computer Science, Gwangju Institute of Science and Technology
123 Chumdangwagi-ro, Buk-gu, Gwangju 61005, South Korea
(Submitted on May 5, 2023; bham@gist.ac.kr)

**Abstract**
Bell inequality violation is a quantitative measurement tool for quantum entanglement. Quantum entanglement is the heart of quantum information science, in which the resulting nonlocal correlation between remotely separated photons shows a unique property of quantum mechanics. Here, the role of coincidence detection is coherently investigated for the nonlocal correlation in a simple polarization-basis selective non-interferometric system using entangled photon pairs (Phys. Rev. A **73**, 012316 (2006)). The resulting nonlocal quantum feature between two independent local polarizers is coherently derived for the joint-parameter relation of the inseparable intensity product. The resulting coherence solution based on the wave nature of quantum mechanics is thus understood as a deterministic process via coincidence detection-caused measurement modification.

**Introduction**
Quantum entanglement is the heart of quantum information science which is known to be impossible by any classical means [1-4]. The definition of 'classical' implies for incoherent and individual particles [2,5]. Bell inequality violation is a test tool for quantitative measurement of quantum entanglement with respect to the local hidden variable theory [5]. Nonlocal correlation between entangled photons shows a mysterious phenomenon known as the EPR paradox [6]. The EPR paradox is for the violation of local realism between space-like separated parties [7]. A common method of entangled photon generation is to use a spontaneous parametric down-conversion (SPDC) process [8,9]. The nonlocal correlation has also been investigated in the delayed-choice quantum eraser [10-12]. Here, coincidence measurements of entangled photon pairs are investigated according to the wave nature of a photon in a noninterfering Sagnac interferometer-based quantum eraser [13] to derive a coherence solution of the quantum feature. For this coherence solution, coincidence measurements play a critical role to induce second-order quantum superposition between paired photons, as in the first-order quantum superposition of a single photon [14].

In 1978, Wheeler proposed a thought experiment of a delayed choice, in which post-measurements can reverse a predetermined photon characteristic [15]. Since then, many experimental proofs have been followed to demonstrate the wave-particle duality [16-20]. The predetermined particle nature of a photon in an orthogonally polarized MZI can be swapped in a time-reversed manner via post-measurements through polarization basis controls [12,13,16]. The post-measurements are for swapping between the wave and particle natures of a photon. In the present study, the nonlocal correlation observed in ref. [13] is coherently interpreted for a delayed-choice quantum eraser to understand the origin of quantum features. For this, an entangled photon pair is coherently analyzed for the polarization-basis projection on a rotated polarizer, and a coherence solution of the nonlocal quantum feature is derived for a joint-parameter relation of local polarizers. The resulting coherence solution of the nonlocal quantum feature shows a perfectly deterministic local property of polarizers. This determinacy is different from conventional coherence optics due to the coincidence detection-caused measurement modifications.

**Results**
Figure 1(a) shows the schematic of a noninterfering Sagnac interferometer based on a quantum eraser scheme [13]. A pump photon comes from the right of the polarizing beam splitter (PBS) through a 22.5-degree rotated half-wave plate (HWP) (not shown). A PBS-split pump photon excites a type-II phase-matched entangled photon pair into either the forward ('2') or backward ('1') direction of the $\chi^{(2)}$ nonlinear crystal. The oppositely detuned (symmetric) frequency relation in each photon pair is shown in Fig. 1(b), resulting in a frequency-correlated photon pair. By the $\chi^{(2)}$ nonlinear optics, a phase-matching condition is satisfied among the pump photon and the resulting entangled photon pair [13,21]. Each pair of entangled photons is deterministically split into both output ports of the PBS depending on their polarization bases. Although the entangled photon pairs from the nonlinear crystal are generated indistinguishably with no which-way information according to the split pump photons by the PBS, the output photons from the PBS turn out to be distinguishable with perfect witch-way information, resulting in no local fringes. Unlike a backward pump scheme in refs. [12,21], resulting in strong phase dependency between superposed photons, the generated entangled-photon pairs '1' and '2' is immune to the phase control $\varphi$ [13]: Due to the phase matching condition, the $\varphi$-controlled pump photon incident from the '1' direction affects only the photon pair exiting along the '2' direction. Instead, the $\varphi$-uncontrolled pump photon-generated photon pair along the '1' direction should be affected by the same $\varphi$, resulting in $\varphi$ cancellation between photon pairs in both directions.

Because of the locked phase between entangled photons by the phase matching of SPDC nonlinear optics [8,9], each pair of PBS output photons denoted by the same color in Fig. 1(a) is phase coherent. This is the motivation of the present coherence approach. However, such phase coherence cannot be applied to different pairs as denoted by different colors due to random detuning [13,22]. In addition to the orthogonal polarization bases, thus, the local interference condition is not satisfied for all measured photons due to $\delta f_{jk} \neq 0$ [23]. Due to a fixed phase between entangled photons in each pair [24], the coincidence detection between local detectors $D_s$ and $D_i$ should be coherent [13], where the random phase between pairs does not affect the results. Recently, such a fixed phase relationship between the signal and idler photons in Fig. 1(b) has been analyzed for $\pi/2$ [24], where this fixed phase relation is the bedrock of the nonlocal fringe between local detectors (analyzed below) [13].

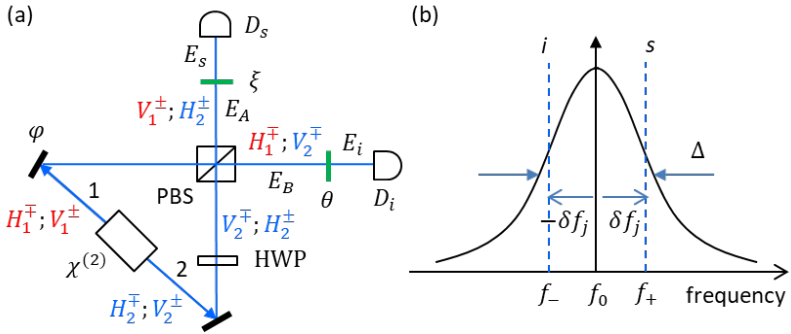

**Fig. 1.** Schematics of quantum correlation (see ref. [13]). (a) Quantum model based on $\chi^{(2)}$ SPDC. (b) SPDC-generated photon spectrum. (d) Coherence model based on coherent photons. A1/A2: synchronized acousto-optic modulators, $D_s$/$D_i$: single-photon detectors, HWP: 22.5°-rotated half-wave plate, L: laser, BS: nonpolarizing 50/50 beam splitter, PBS: polarizing beam splitter, $\xi$, and $\theta$ are polarizer's rotation angles. In (c), a heterodyne detection method is adapted for coincidence measurements.

In Fig. 1(a), the photon pair indicated by the same subscript (or same color) must be coherent with each other by the $\chi^{(2)}$ phase-matching condition [8,9], whose frequencies are oppositely detuned, as shown in Fig. 1(b). The symmetric detuning $\pm\delta f_j$ of the $j^{th}$ photon pair across the center frequency $f_0$ is represented by '$\pm$' in the superscript of the polarization bases. Here, a simultaneous generation of both photon pairs in opposite directions is negligible [8,9,13,21].

In Fig. 1(b), the $j^{th}$ paired photons are oppositely detuned by $\pm\delta f_j$ from the center frequency $f_0$ to satisfy the energy conservation law in a degenerate SPDC process, where the center frequency is half of the pump frequency. The frequency stability between paired photons strongly depends on the pump linewidth [8,9]. Due to the phase-matching condition of SPDC, the coherence relation is strictly satisfied among the pump, signal, and idler photons within the geometry [21]. In addition to the fixed phase, the symmetric frequency detuning between entangled photons in each pair plays an important role in the bandwidth-independent joint-parameter relation in nonlocal correlation (analyzed below).

In Fig. 1(a), the output photons out of the noninterfering Sagnac interferometer are coherently represented as:

$$E_A = \frac{E_0}{\sqrt{2}} e^{i\varphi'} \left( -V_1^{\pm} e^{i\delta f_j t} + H_2^{\pm} e^{i\delta f_k t} \right), \quad (1)$$

$$E_B = \frac{iE_0}{\sqrt{2}} e^{i\varphi'} \left( H_1^{\mp} e^{-i\delta f_j t} + V_2^{\mp} e^{-i\delta f_k t} \right), \quad (2)$$

where $E_0$ is the amplitude of a single photon. In Eqs. (1) and (2), the single photon is represented for a harmonic oscillator with coherence optics according to the wave nature of a photon, where $\varphi'$ is for the traveling wave including both phase $\varphi$ and the global phase. In Eqs. (1) and (2), the fixed relative phase $\pi/2$ between the signal and idler photons is considered [24]. The opposite-sign notation between the same polarization basis (see the superscript) is due to the half-wave plate (HWP) applied on path 2, resulting in polarization-basis swapping. Thus, the orthogonal polarization bases between paths "1" and "2" must have the same detuning in Eqs. (1) and (2). Due to the orthogonal polarization bases on PBS, the corresponding mean intensities of Eqs. (1) and (2) are $\langle I_A \rangle = \langle I_B \rangle = \langle I_0 \rangle$, resulting in distinguishable photon characteristics.

With the inserted polarizers $(\xi, \theta)$ in both output paths of the PBS, Eqs. (1) and (2) are rewritten for polarization projections of the output photons onto the rotated polarization angles $\xi$ and $\theta$:

$$E_s = \frac{E_0}{\sqrt{2}} e^{i\varphi'} \left( -V_1^{\pm} \sin\xi e^{i\delta f_j t_1} + H_2^{\pm} \cos\xi e^{i\delta f_k t_2} \right), \quad (3)$$

$$E_i = \frac{iE_0}{\sqrt{2}} e^{i\varphi'} \left( H_1^{\mp} \cos\theta e^{-i\delta f_j t_1} + V_2^{\mp} \sin\theta e^{-i\delta f_k t_2} \right), \quad (4)$$

where the inclusion of the polarization bases $(H_{1,2}; V_{1,2})$ is just to indicate the photon's origin for further analysis of coincidence detection below. The corresponding intensities are as follows:

$$I_s = \frac{I_0}{2} \left( 1 - \sin 2\xi \sin(\delta_{jk}) \right), \quad (5)$$

$$I_i = \frac{I_0}{2} \left( 1 + \sin 2\theta \sin(\delta_{jk}) \right), \quad (6)$$

where $\delta_{jk} = \delta f_j t_1 - \delta f_k t_2$, and $H_j V_k$ basis-product term is allowed by the projection-driven common polarization basis, resulting in the indistinguishable photon characteristics. Even though the geometrical phase lock between the forward and backward photon pairs can be established by the $\chi^{(2)}$ phase matching, the frequency detuning-caused phase difference, i.e., $\delta_{jk}$ is for the photon ensemble of all pairs in Fig. 1(b), resulting in a uniform intensity, too [24]: $\langle I_s \rangle = \langle I_i \rangle = \frac{\langle I_0 \rangle}{2}$ due to $\langle \sin(\delta_{jk}) \rangle = 0$ [23]. Here, the half-cut in intensity is because of the 50% photon loss by the polarizer. By the way, for the backward pump scheme in ref. [12], however, the photon phase of the forward pair is influenced by the backward one, resulting in pair-to-pair phase matching [20,22]. In this case, the local randomness is partially violated [12].

From Eqs. (3) and (4), the coincidence detection between detectors $D_s$ and $D_i$ is for the same colored-photon (orthogonal polarization) pairs, where different colored-photon pairs are selectively discarded by the definition of coincidence:

$$\langle R_{si}(\tau = 0) \rangle = \langle (E_s E_i)(E_s E_i)^* \rangle$$
$$= \frac{\langle I_0^2 \rangle}{4} \langle \left( -V_1^{\pm} H_1^{\mp} \cos\theta \sin\xi + H_2^{\pm} V_2^{\mp} \sin\theta \cos\xi \right)(cc) \rangle$$
$$= \frac{\langle I_0^2 \rangle}{4} \langle \sin^2(\theta - \xi) \rangle. \quad (7)$$

Interestingly, Eq. (7) shows invariance on both $\varphi$ and $\Delta$. Here, the $\delta f_j$ immunity is due to the opposite detuning relation by the phase-matching condition of SPDC, where $V_1^{\pm} H_1^{\mp}$ notation is just to indicate the origin of interacting photons. Due to the $\varphi$ invariance, the coincidence detection applies only after PBS. Due to the same pair interactions, the mean value of the coincidence detections is for the ensemble average such as in the Hong-Ou-Mandel effect showing the ensemble decoherence in the order of $\Delta^{-1}$ [24,25]. This coherence solution of the nonlocal quantum feature in Eq. (7) is equivalent to the observations in ref. [13], where coincidence detection plays a critical role in the joint parameter relation for the product-basis selections, resulting in the second-order quantum superposition. This is the extended indistinguishable photon characteristics between product bases in a two-photon correlation.

Mathematically, the Bell parameter S for the $\langle R_{si}(0)\rangle$ in Eq. (7) is $2\sqrt{2}$ due to the perfect sinusoidal modulations [5,13,26,27]. Due to the phase correlation by SPDC, the space-like separation between two local detectors won't affect the measurement results because the coherence time between paired photons is much longer than the effective coherence $\Delta^{-1}$. Thus, the nonlocal quantum feature satisfying the inseparable intensity product is successfully analyzed for Fig. 1 using coincidence detection-caused selective measurements. From this analysis, it can be said that the coherence-based polarization correlation between paired photons is the bedrock of the spooky action at a distance as Einstein argued for the hidden variable [2].

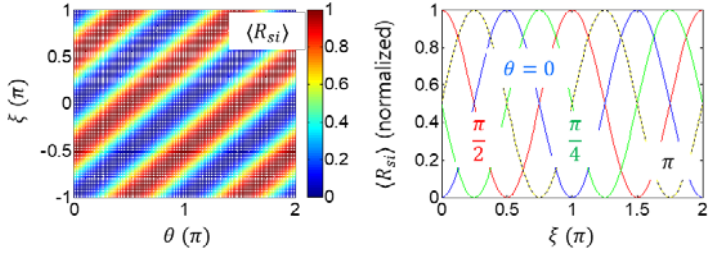

**Fig. 2.** Numerical calculations for Eq. (8).

Figure 2 shows the numerical calculations for Eq. (7), resulting in the same nonlocal quantum feature observed in ref. [13]. The joint-parameter relation is shown in the right panel. As mentioned above, the Bell parameter S violates the Bell inequality for hidden variable theory [26]. Thus, the coherence solution derived for Fig. 1 is successfully demonstrated for the entangled photons via coincidence detection-caused measurement modifications. The origin of the inseparable basis products in Eq. (7) is the second-order quantum superposition between product bases resulting from the coincidence detection. Without coincidence detection, the two photon correlation $\langle R_{si}(\tau)\rangle$ results in the classical limit.

**Conclusion**
The origin of the nonlocal quantum feature was investigated using SPDC-generated entangled photon pairs in a noninterfering Sagnac-based quantum eraser scheme [13]. For this, firstly, the phase-matched photon pairs whose polarizations are orthogonal were prepared inside the Sagnac interferometer using a coherence approach. Secondly, coincidence measurements between local detectors were coherently analyzed for polarization projections of a pair of independent polarizers. Thirdly, the nonlocal quantum feature of Bell inequality violation was coherently derived from the coincidence detection-caused measurement modification, resulting in an inseparable joint-parameter relation. As a result, the coherence solution of the path-length- and bandwidth-independent nonlocal correlation was derived analytically and numerically confirmed for the same quantum features observed in ref. [13]. Due to the coherence feature of the paired photons for a fixed phase relation, thus, the nonlocality is understood as a deterministic process of coherence optics for the phase matching of the SPDC process via coincidence detection-modified product-basis superposition. This coherence interpretation enriches our understanding of otherwise mysterious quantum correlation.

**Acknowledgment**
BSH acknowledges the helpful discussions with Prof. T. Kim at Seoul National University, South Korea.

**Funding**

his research was supported by the MSIT(Ministry of Science and ICT), Korea, under the ITRC(Information Technology Research Center) support program(IITP-2023-2021-0-01810) supervised by the IITP(Institute for Information & Communications Technology Planning & Evaluation). BSH also acknowledges that this work was partially supported by GIST via GRI-2023.